\documentclass[aps,pre,notitlepage]{revtex4-1}
\usepackage{amsmath,amsfonts,mathtools,amssymb}
\usepackage[colorlinks=true,linkcolor=blue,citecolor=red,bookmarksnumbered=true]{hyperref}
\usepackage{times,mathpazo,charter,graphicx,bm,enumitem}
\usepackage{color}

\makeatletter
\renewcommand\section{\@startsection {section}{1}{\z@}%
  {-4.6ex \@plus -0.4ex \@minus -0.2ex}{2.4ex \@plus0.2ex}%
  {\normalfont\bfseries}}
\renewcommand\subsection{\@startsection{subsection}{2}{\z@}%
  {-3.2ex\@plus -0.2ex \@minus -0.2ex}{1.8ex \@plus0.2ex}%
  {\normalsize\it}}
\renewcommand\paragraph{\@startsection{paragraph}{4}{\z@}%
  {2.0ex \@plus 0.2ex \@minus 0.2ex}{-1em}%
  {\normalfont\normalsize\bfseries}}
\def\@maketitle{\null\vskip2.4em%
  \begin{center} \let\footnote\thanks \vskip3.4em%
  {\Large\@title \par}\vskip2.4em%
  {\lineskip 0.5em\begin{tabular}[t]{c}\@author\end{tabular}\par}\vskip1em%
  {\@date}%
  \end{center}%
  \par\vskip3.4em}
\makeatother

\allowdisplaybreaks

\let\^=\hat
\def\bmu{\boldsymbol\mu}
\def\bPhi{\boldsymbol\Phi}
\def\@#1{{\mathbf{#1}}}
\def\diag{\mathop{\rm diag}\nolimits}
\def\Real{\mathbb{R}}
\def\Integer{\mathbb{Z}}
\def\Complex{\mathbb{C}}
\def\e{\mathrm{e}}

\def\[{\begin{equation}}
\def\]{\end{equation}}
\def\be{\begin{equation}}
\def\ee{\end{equation}}
\def\bse{\begin{subequations}}
\def\ese{\end{subequations}}

\def\idots{\hbox{\reflectbox{$\ddots$}}}

\begin{document}
\title{Discrete and continuous coupled nonlinear integrable systems via the dressing method}
\author{Gino Biondini}
\author{Qiao Wang}
\affiliation{Department of Mathematics, State University of New York at Buffalo, Buffalo, New York 14260, USA}

\begin{abstract}
A discrete analogue of the dressing method is presented and used to derive integrable nonlinear evolution equations, including two
infinite families of novel continuous and discrete coupled integrable systems of equations of nonlinear Schr\"odinger type.
First, a demonstration is given of how discrete nonlinear integrable equations can be derived starting from their linear counterparts.
Then, starting from two uncoupled, discrete one-directional linear wave equations, an appropriate matrix Riemann-Hilbert problem is constructed, 
and a discrete matrix nonlinear Schr\"odinger system of equations is derived, together with its Lax pair.
The corresponding compatible vector reductions admitted by these systems are also discussed, as well as their continuum limits.
Finally, by increasing the size of the problem,
three-component discrete and continuous integrable discrete systems are derived,
as well as their generalizations to systems with an arbitrary number of components.
\end{abstract}

\maketitle

\section{Introduction}
\label{s:intro}

Nonlinear integrable systems play an important role in mathematical physics.
From a theoretical point of view, they possess a rich mathematical structure, 
and in many cases they are exactly solvable. 
From a concrete point of view, 
they often arise as the governing equations in many physical applications.

The theory of infinite-dimensional continuous integrable system has been extensively developed and studied in the last fifty years
(e.g., see \cite{AC1991,ablowitzsegur1981,BBEIM1994,FT1987,GH1990,NMPZ1984,TrogdonOlver,Yang}).
The effort to extend our knowledge of continuous integrable systems to discrete nonlinear systems has also been an ongoing theme 
since the works of Ablowitz and Ladik~\cite{ablowitzladik}, Toda~\cite{toda}, Flaschka~\cite{flaschka,flaschka2} and many others,
continuing to the present day
(e.g. see~\cite{ablowitzhalburdherbst,ablowitzprinaritrubatch,bobenkoseiler,bruschi1,bruschi2,doliwa,doliwa2,%
goktashereman,grammaticos,halburd,hietarintajoshinijhoff,kajiwarasatsuma,kakei,leviwinternitz,ohta,ohtahirota,santini} and references therein). 
In particular, a key question is whether there exist one or more integrable discrete analogues to a given continous system.
Another key question is whether a given linear system admits a completely integrable nonlinearization. 

There are several approaches to obtain nonlinear integrable systems starting from their linear counterpart.
One approach is the Ablowitz-Kaup-Newell-Segur method~\cite{akns1974,ablowitzsegur1981}, which starts from the linear dispersion relation. 
Another approach is the so-called dressing method, first introduced in~\cite{zakharovshabat1,zakharovshabat2}.
The key ingredient in this approach is the formulation of a suitable Riemann-Hilbert problem (RHP) \cite{ablowitzfokas,Gakhov,TrogdonOlver},
which relates the limiting values of a scalar or matrix sectionally analytic function across a given curve.
In particular, it was shown in~\cite{fokas2008,fokaszakharov} that the dressing method can be used to derive nonlinear systems staring 
from linear ones 
by appropriately modifying the RHP, see also~\cite{pinotsis2007} for further details. 
The technique was then recently used by the authors to derive novel systems of resonant wave interactions \cite{BiondiniWang2015}.

The purpose of this work is to show that similar techniques can also be successfully applied to discrete integrable systems. 
In the process we derive infinite families of continuous and discrete coupled systems of interactions of nonlinear Schr\"{o}dinger (NLS) type, 
which are novel to the best of our knowledge.
Importantly, the simplest among these systems 
[most notably, \eqref{e:d2novel}, \eqref{e:c2novel}, \eqref{e:d3novel} and \eqref{e:c3novel}]
are continuous and discrete analogues of the partial differential equations (PDEs) that describe certain spinor systems~\cite{spinor1,spinor2,spinor3} in Bose-Einstein condensates \cite{Kevrekidis,Pitaevskii},
which have attracted considerable attention in recent years,
and are therefore likely to be physically relevant on their own.

This work is organized as follows.
In section~\ref{s:dLS} we introduce the discrete analogue of the dressing method and show how one can recover a discrete linear evolution equation 
and its Lax pair from the knowledge of just its associated RHP.
In section~\ref{s:dNLS} we then ``nonlinearize'' the method and use it to derive the integrable discrete nonlinear Schr\"{o}dinger equation together with its Lax pair
starting from the RHP of its linear counterpart.
In section~\ref{s:dmatrixNLS} we generalize the method to coupled systems, 
and we present the derivation of a discrete matrix NLS equation.
In section~\ref{s:dcoupledNLS} we discuss the reduction of the matrix system to various two-component systems
such as the discrete analogue of the Manakov system and of a spinor system,
both of which have been intensely studied for many years~\cite{ablowitzprinaritrubatch,spinor1,manakov,NMPZ1984,spinor2,spinor3}.
Then in section~\ref{s:novel} we increase the size of the problem and we derive two infinite families of novel continuous and discrete 
coupled integrable systems of NLS type with arbitrary numbers of components. 
Section~\ref{s:remarks} offers some final remarks.
A few results are relegated to the appendices.
Specifically, in appendix~\ref{s:cNLS} we review the derivation of the NLS equation through the dressing method, 
which was also discussed in~\cite{pinotsis2007},
and in appendix~\ref{s:ccoupledNLS} we derive the matrix NLS equation together with its Lax pair,
which serves to illustrate the main differences between the continuous and the discrete versions of the technique.
Finally, in appendix~\ref{s:mKdV}, using the one-directional wave equation,
we derive a discrete integrable system which admits a reduction to a discrete modified Korteweg-deVries (KdV) equation,
demonstrating that the method is 
not limited to equations of NLS type,
and has instead a broader scope of applicability.


\section{Lax pair of discrete linear evolution equations via dressing}
\label{s:dLS}

We begin to present the dressing method for discrete systems
by showing how the discrete linear Schr\"odinger equation and its Lax pair can be recovered from the knowledge of 
the associated RHP,
which will serve to introduce some of the relevant notation.

Consider the discrete linear Schr\"odinger equation, i.e., the differential-difference system
\[
i\dot q_n + (q_{n+1} -2q_n + q_{n-1}) = 0\,,
\label{e:dLS}
\]
where the dot denotes temporal derivative.  
One can also write \eqref{e:dLS} as $i\dot q_n + (\e^{\partial} - 2 + \e^{-\partial})\,q_n = 0$
by introducing the shift operator
$e^\partial$ defined as $e^\partial f_n = f_{n+1}$.
which we will use throughout this work.
Equation~\eqref{e:dLS} is a discrete analogue of the linear Schr\"odinger equation $iq_t + q_{xx} = 0$ since it reduces to it
when the terms in parenthesis (which identify the second-order central difference throughout this work)
are divided by $h^2$
and one takes the limit $h\to0$ with $x_n= nh$ (the shift operator becoming $\e^{h\partial}$).

It was shown in~\cite{BiondiniHwang2008} that \eqref{e:dLS} is associated to the scalar RHP defined by the jump condition
\vspace*{-0.4ex}
\bse
\label{e:dLS-RHP}
\[
\phi_n^+(t,z) - \phi_n^-(t,z) = z^{n-1}e^{-i\omega_0(z)t}\^q(z), \qquad |z| = 1\,,
\label{e:dLS-jump}
\] 
for the sectionally analytic function
\vspace*{-0.4ex}
\[
\phi_n(t,z) = \begin{cases} \phi_n^+(t,z), & |z|<1, \,\\ \phi_n^-(t,z), & |z|>1,\end{cases}
\label{e:dLS-phidef}
\]
where $\omega_0(z)$ denotes the linear dispersion relation of~\eqref{e:dLS}, i.e., 
\vspace*{-0.4ex}
\[
\omega_0(z) = 2 - (z + 1/z)\,,
\label{e:dLS-lineardisp}
\]
\ese
and $\^q(z)$ is the discrete Fourier transform of $q_n(0)$, 
namely $\^q(z) = \sum_{n\in\Integer} q_n(0)/z^n$.
Note that here the jump condition is given over the unit circle in the complex plane (i.e. $|z|=1$, 
unlike the continuous case, where it is given across the real $k$-axis \cite{BiondiniWang2015}).
As a result, one has two separate asymptotic behaviors: for $\phi_n^+(t,z)$ 
 as $z\to0$  
and for $\phi_n^-(t,z)$ 
 as $z\to0$
(instead of the single limit point $k\to\infty$).
In \cite{BiondiniHwang2008}, the jump condition~\eqref{e:dLS-RHP} was obtained via spectral analysis of the Lax pair associated with~\eqref{e:dLS}, namely
\vspace*{-1ex}
\bse
\label{e:dLS-LP}
\begin{gather}
   \phi_{n+1} - z\phi_n = q_n\,,  \\
   \dot \phi_n + i\omega_0(z)\phi_n = iq_n - iq_{n-1}/z\,.
\end{gather}
\ese
In particular, the analysis of the scattering problem in~\cite{BiondiniHwang2008} shows that 
\vspace*{-1ex}
\bse
\label{e:dLS-phiasymp}
\begin{gather} 
\phi_n^+(t,z)=O(z) \quad z \to 0,	\\
\phi_n^-(t,z) = O(1/z) \quad z \to \infty.
\end{gather}
\ese
Futhermore, it is straightforward to show that the compatibility condition of~\eqref{e:dLS-LP} 
[namely, requiring that $e^\partial(\partial_t(\phi_n)) = \partial_t(e^\partial(\phi_n))$] 
yields the discrete linear Schr\"{o}dinger equation~\eqref{e:dLS}.
Importantly, however, we next show that the jump condition~\eqref{e:dLS-jump} together with the asymptotic behavior~\eqref{e:dLS-phiasymp}
are sufficient to show that $\phi_n(t,z)$ satisfies the Lax pair~\eqref{e:dLS-LP} and therefore to recover the discrete linear Schrodinger equation~\eqref{e:dLS}.
This is the key that will allow one to ``nonlinearize'' the procedure and obtain discrete nonlinear integrable systems.

The main idea of the dressing method for discrete systems, 
similarly to that for continuous systems~\cite{fokas2008}, 
is the following:
Starting from an appropriate RHP, one constructs two linear operators $L$ and $N$ such that
(i)~$L\phi_n$ and $N\phi_n$ 
have no jump across the unit circle, 
and (ii)~$L\phi_n$ and $N\phi_n$ vanish as $z\to\infty$ and 
 
remain bouded as $z\to0$, i.e., 
\bse
\label{e:dLS-LNasymp}
\begin{gather} 
L \phi_n^+(t,z) =  O(1),  \quad z \to 0,\qquad
L \phi_n^-(t,z) = O(1/z), \quad z \to \infty,
\label{e:dLS-Lasymp}
\\
N \phi_n^+(t,z) =  O(1),  \quad z \to 0,\qquad
N \phi_n^-(t,z) = O(1/z), \quad z \to \infty.
\label{e:dLS-Nasymp}
\end{gather}
\ese
Then, under the assumption that the RHP has a unique solution, one can conclude that $L\phi_n$ and $N \phi_n$ are identically zero, 
which yields the Lax pair associated with the RHP. 
Next we show in detail how this approach can be carried out.  

We first derive the scattering problem.  We begin by defining 
\[
L\phi_n = (e^\partial-z)\phi_n - q_n\,.
\label{e:dLS-Ldef}
\]
The $z$ dependence and the shift operator are dictated by the requirement that $L\phi_n$ 
has no jump across the unit circle.
The function $q_n(t)$, which is independent of $z$, is at this point arbitrary, but is required in order for $L\phi_n$ to satisfy 
the asymptotic behavior \eqref{e:dLS-Lasymp}.
It is straightforward to see that $L\phi_n^+ = L\phi_n^-$.
Next, using~\eqref{e:dLS-phiasymp} we expand $\phi_n$ as
\bse
\label{e:dLS-phiexpand}
\begin{gather}
\phi_n^+(t,z) = \phi_n^{(0,+)} + \phi_n^{(1,+)}z + \phi_n^{(2,+)}z^2 + O(z^3), \quad z\to 0,  \\
\phi_n^-(t,z) = \phi_n^{(0,-)} + \phi_n^{(1,-)}/z + \phi_n^{(2,-)}/z^2 + O(1/z^3), \quad z\to \infty.
\end{gather}
\ese
Substituting~\eqref{e:dLS-phiexpand} into~\eqref{e:dLS-Ldef} 
one can verify that the asymptotic condition~\eqref{e:dLS-Lasymp} is satisfied.
We have thus obtained the fist half of~\eqref{e:dLS-LP}.
Moreover, if $\smash{\phi_n^{(0,-)}} = 0$,  at $O(1/z)$, $O(1/z^2)$ and $O(1/z^2)$ as $z\to\infty$ 
one finds, respectively,
$\smash{\phi_n^{(1,-)}} = -q_n$,
$\smash{\phi_n^{(2,-)}} = -q_{n+1}$ and 
$\smash{\phi_n^{(3,-)}} = - q_{n+2}$.

Now we derive the time dependence part of the Lax pair.
Similarly as before, to get the second half of~\eqref{e:dLS-LP}, we look for an operator in the form
\[
N\phi_n = \dot \phi_n + i\omega_0(z)\phi_n - [A_n(t)z +B_n(t)/z + C_n(t)].
\label{e:dLS-Ndef}
\]
Again, the time derivative and dependence on $\omega_0(z)$ is dictated by the requirement that $N\phi_n$ 
has no jump across the unit circle,
and $A_n(t),B_n(t)$ and $C_n(t)$ are functions of $t$ that are independent of $z$.
We have
\[
N\phi_n^+ = N\phi_n^-\,,
\]
which again can be easily checked by direct substitution.
The terms in the square bracket in \eqref{e:dLS-Ndef} must be determined by requiring that  $N\phi_n$ satisfies the asymptotic behavior~\eqref{e:dLS-Nasymp} 
[i.e., vanishes as $z\to0$ and as $z\to\infty$].
We thus substitute~\eqref{e:dLS-phiexpand} into~\eqref{e:dLS-Ndef}, impose~\eqref{e:dLS-Nasymp} and solve for
$A_n(t),B_n(t)$ and $C_n(t)$. 
Specifically, 
requiring that the term $O(z)$ vanishes as $z\to\infty$
yields $A_n = 0$,
and that at $O(1)$ yields $C_n = iq_n$.
The term at $O(1/z)$ yields 
$B_n = i(i\dot q_n + q_{n+1}- 2q_n)$.
Finally, the term at $O(1/z^2)$ yields Eq.~\eqref{e:dLS}. 
Backsubstituting, we then obtain $B_n = -iq_{n-1}$ and thus
recover the second half of~\eqref{e:dLS-LP}.

\section{Integrable discrete nonlinear Schr\"{o}dinger equation via dressing}
\label{s:dNLS}

Now we show how one can properly ``dress'' (i.e., modify) the RHP~\eqref{e:dLS-RHP} and derive the integrable discrete NLS equation,
together with its associated Lax pair,
starting from its linear counterpart, \eqref{e:dLS}.
Recall that 
the integrable discrete NLS equation, also known as the Ablowitz-Ladik system~\cite{ablowitzladik}, is the differential-difference system
\[
i\dot{q}_n + (q_{n+1} - 2q_n + q_{n-1}) - \nu |q_n|^2 (q_{n+1} + q_{n-1}) = 0\,,
\label{e:dNLS-AL}
\]
with $\nu = \pm 1$ denoting the defocusing and focusing cases, respectively.


Similarly to the continuous case (see appendix~\ref{s:cNLS} for details), we can trivially change the above scalar RHP~\eqref{e:dLS-RHP} into a matrix RHP by introducing
the matrix
\[
M_n(t,z) = \begin{pmatrix} 1&\phi_n(t,z) \\0&1\end{pmatrix},
\label{e:dLS-matrixdef}
\] 
obtaining the jump condition
\[
M_n^+(t,z) = M_n^-(t,z)V_n(t,z), \qquad |z| = 1\,,
\label{e:dLS-matrixjump}
\]
with jump matrix
\[
V_n(t,z) =  \begin{pmatrix} 1&z^{n-1}e^{-i\omega_0(z)t}f(z) \\0&1\end{pmatrix}.
\label{e:dLS-jumpmatrix}
\]
Note that $M_n(t,z)$ does not satisfy the same asymptotic behavior as $\phi_n(t,z)$ as $z\to0$ and $z\to\infty$.
Instead, 
$M_n^+(t,z) = I + O(z)$ as $z\to 0$ and 
$M_n^-(t,z) = I + O(1/z)$ as $z\to \infty$.
Since the goal now is not to solve an initial value problem, but rather to obtain a novel evolution equation,
the function $\^q(z)$ in the jump condition was replaced with an arbitrary function $f(z)$ defined on $|z|=1$.

The ``nonlinearization'' step consists in replacing the jump matrix $V_n(t,z)$ above with the new matrix
\[
R_n(t,z) = V_n^\dagger(t,z^2)V_n(t,z^2) 
= \^{Z}^{n-1}e^{-i\omega(z)t\^\sigma_3}S(z^2),
\label{e:dNLS-Rdef}
\]
where 
$V^\dagger = (V^*)^T$ denotes the matrix adjoint, 
with the asterisk $*$ and superscript $T$ denoting respectively complex conjugation and matrix transpose,
\[
Z = \begin{pmatrix} z&0 \\0&1/z\end{pmatrix}, \qquad 
S(z) = \begin{pmatrix} 1&&f(z) \\f^*(z)&&1+|f(z)|^2\end{pmatrix}
\label{e:dNLS-Zdef}
\]
and $\omega(z) = -(z - 1/z)^2/2.$
Equation~\eqref{e:dNLS-Rdef} is the discrete analogue of the nonlinearization process in the continuous case (e.g., see \cite{fokas2008,pinotsis2007}).
Note also that $\omega(z) = \frac12\omega_0(z^2)$.
The change $z\mapsto z^2$ is the discrete analogue of the rescaling $k\mapsto 2k$ when recovering 
the direct and inverse Fourier transform from the linear limit of the inverse scattering transform \cite{BiondiniHwang2008}.
This can also be easily seen by recalling that the correspondence between the spectral variables in the discrete and continuous case is  $z = e^{i k h}$.

Above and throughout this work, we use the notations $\^Z A$ and 
$e^{s\^\sigma_3} A$ to denote the similarity transformations $\^Z A = ZAZ^{-1}$ 
and $e^{s\^\sigma_3} A = e^{s\sigma_3} A\, e^{-s\sigma_3}$ 
for any matrix $A$ and scalar $s$.
Note that in \eqref{e:dNLS-Rdef} we used the fact that $z^* = 1/z$ and $\omega(z) = \omega^*(z) $ on $|z| = 1$.
We therefore consider the following modified RHP for the matrix
\[
M_n(t,z)=
\begin{cases}
    M_n^+(t,z) \qquad |z|<1,  \\
    M_n^-(t,z) \qquad |z|>1,  \\
  \end{cases}
\label{e:dNLS-Mdef} 
\]
with 
asymptotic behavior as $z\to0$ and $z\to\infty$ given respectively by%
\bse
\label{e:dNLS-Masymp0}
\begin{gather}
\label{e:dNLS-M+asymp}
M_n^+(t,z) = M_n^{(0,+)} + O(z) \qquad z\to 0,  \\
\label{e:dNLS-M-asymp}
M_n^-(t,z) = M_n^{(0,-)} + O(1/z) \qquad z\to \infty, 
\end{gather}
\ese
with $M_n^{(0,-)} = I$ and $M_n^{(0,+)}$ to be determined,
and the modified jump condition
\[
M_n^+(t,z) = M_n^-(t,z)R_n(t,z),\qquad |z|=1\,.
\label{e:dNLS-jump}
\]
(Note that $M_n^{(0,+)}$ can differ from the identity because 
the normalization as $z\to\infty$ and the jump condition are enough to determine $M_n(t,z)$ completely, 
and therefore 
one does not have the freedom to also specify a normalization condition as $z\to0$.)  
Below we show that any solution $M(t,z)$ to the above RHP satisfies the matrix Lax pair%
\bse
\label{e:dNLS-LP}
\begin{gather}
M_{n+1} - \^{Z}M_n = Q_nM_nZ^{-1},  \\
\dot{M}_n + i\omega(z)[\sigma_3,M_n] = H_nM_n,
\end{gather}
\ese
where $[A,B] = AB - BA$ is the matrix commutator,
with
\[
H_n = i\sigma_3(Q_nZ^{-1} - Q_{n-1}Z - Q_nQ_{n-1}).
\label{e:dNLS-Hdef}
\]
Moreover, the matrix
\[
Q_n(t) = \begin{pmatrix} 0&q_n \\r_n&0\end{pmatrix}
\label{e:dNLS-Qdef}
\]
satisfies the matrix discrete NLS equation
\[
i\sigma_3\dot{Q}_n + (Q_{n+1} - 2Q_{n} + Q_{n-1}) - Q_n^2(Q_{n+1} + Q_{n-1}) = 0,
\label{e:dNLS}
\] 
which is the compatibility condition of~\eqref{e:dNLS-LP}.
Under the symmetry reduction $r_n=\nu q_n^*$ with $\nu=\pm1$,
\eqref{e:dNLS} then yields the Ablowitz-Ladik system~\eqref{e:dNLS-AL}.
Note that the above Lax pair is equivalent to the one satisfied by the Jost eigenfunctions \cite{ablowitzprinaritrubatch}, as can be seen 
by letting $v_n(t,z) = M_n(t,z)\,Z^n\,\exp[-i\omega(z)t\sigma_3]$.

The derivation of the matrix Lax pair proceeds in a similar way as in the linear case.
The key difference is that the jump condition~\eqref{e:dNLS-jump} is homogeneous.
Therefore,
we aim to construct two operators $L$ and $N$ such that
$L M_n$ and $N M_n$ both satisfy the same jump condition as $M_n$ on $|z|=1$, and
$L M_n$ and $N M_n$ are both $O(z)$ as $z\to0$ and $O(1/z)$ as $z\to\infty$.
Then, under the assumption that the RHP has a unique solution, one can conclude that $L M_n$ and $N M_n$ are identically zero, 
which yields the desired Lax pair. 

To derive the scattering problem,
similarly to the case of the discrete linear Schr\"o\-dinger equation, we define a linear operator $L$ in the form
\[
LM_n = (e^\partial - \^{Z})M_n - Q_nM_nZ^{-1}\,.
\label{e:dNLS-Ldef}
\]
The proof that $LM_n$ satisfy the same jump condition as $M_n$ for the discrete case is slightly different from the continuous case, 
and therefore we provide it explicitly here. 
To show that
\[
LM_n^+Z = (LM_n^-Z)R_n\,,
\label{e:dNLS-Ljump}
\]
note that the left-hand side of~\eqref{e:dNLS-Ljump} equals 
\[
LM_n^+Z 
= L(M_n^-R_n)Z 
= M_{n+1}^-R_{n+1}Z - ZM_n^-R_n - Q_nM_n^-R_n,
\nonumber
\]
whereas the right-hand side of ~\eqref{e:dNLS-Ljump} is 
\[
(LM_n^-Z)R_n 
= M_{n+1}^-ZR_n - ZM_n^-R_n - Q_nM_n^-R_n.
\nonumber
\]
The two sides are then equal, since~\eqref{e:dNLS-Rdef} implies 
\[
R_{n+1}Z = Z^ne^{-i\omega(z)t\^\sigma_3}S(z^2)Z^{1-n} = ZR_n.
\nonumber
\]
Next, using the normalization condition~\eqref{e:dNLS-M+asymp} and~\eqref{e:dNLS-M-asymp} we can write $M_n$ as
\bse
\label{e:dNLS-Masymp}
\begin{gather}
M_n^+(t,z) = M_n^{(0,+)} + M_n^{(1,+)}z + M_n^{(2,+)}z^2 + O(z^3), \quad z\to 0,  \\
M_n^-(t,z) = M_n^{(0,-)} + M_n^{(1,-)}/z + M_n^{(2,-)}/z^2 + O(1/z^3), \quad z\to \infty.
\end{gather}
\ese
Substituting~\eqref{e:dNLS-Masymp} into~\eqref{e:dNLS-Ldef} and enforcing the asymptotic condition~\eqref{e:dLS-LNasymp},
the relevant calculations, while straightforward, are tedious, and were performed using Mathematica. 
In this way one obtains the first half of~\eqref{e:dNLS-LP}.  

To derive the time dependence equation, we now look for a linear operator in the form
\[
NM_n = \dot{M}_n + i\omega(z)\^\sigma_3M_n + (A_nz +B_n/z + C_n)M_n,
\label{e:dNLS-Ndef}
\]
where as before the time derivative and dependence on $\omega(z)$ ensure that $NM_n$ satisfies the jump condition.
and $A_n,B_n$ and $C_n$ are all $2\times2$ matrices independent of $z$, to be determined.
We then have
\[
NM_n^+ = (NM_n^-)R\,.
\]
Next, one substitutes~\eqref{e:dNLS-Masymp} into~\eqref{e:dNLS-Ndef} and solves recursively 
[i.e., first at $O(z^2)$ and $O(1/z^2)$, then at $O(z)$ and $O(1/z)$ and finally at $O(1)$ as $z\to0$ and $z\to\infty$, respectively]
in order to obtain the expressions of the matrices $A_n,B_n$ and $C_n$. 
The result, as announced earlier, is exactly the second half of~\eqref{e:dNLS-LP}.
For brevity,
in this case
we omit the details.

Note that the Lax pair~\eqref{e:dNLS-LP} is of course just a slight reparametrization of the general ``$qr$'' matrix Lax pair of the Ablowitz-Ladik system \cite{ablowitzprinaritrubatch}.
As such, it also admits reductions to other integrable nonlinear evolution equations.  
In particular, taking $r_n = \nu q_{-n}^*$ yields the nonlocal Ablowitz-Ladik system
recently studied in \cite{AM3}.

\section{Discrete matrix NLS systems via dressing}
\label{s:dmatrixNLS}

We now extend the results of the previous section to coupled systems.
Specifically, starting from two uncoupled discrete linear Schr\"{o}dinger equations and their associated RHPs, 
and following similar steps as in the previous sections, we derive a matrix NLS equation together with its Lax pair. 
Then, in the following section, we discuss the reductions that yield the discrete Manakov system and a two-component  reduction of the discrete spinor system.
To avoid confusion, we will use boldface letters to denote $2\times 2$ matrices.


Consider the following two uncoupled sectionally analytic functions,
\bse
\label{e:dManakov-ScalarRHP} 
\[
\phi_n^{(j)}(t,z) = \begin{cases} \phi_n^{(j,+)}(t,z), & |z|<1, \,\\ \phi_n^{(j,-)}(t,z), & |z|>1,\end{cases}
\]
for $j=1,2$, 
which satisfy two uncoupled scalar RHPs defined by the jump conditions
\[
\phi_n^{(j,+)}(t,z) - \phi_n^{(j,-)}(t,z) = z^{n-1}e^{-i\omega_0(z)t}f(z) \qquad |z| = 1, 
\label{e:dManakov-jump} 
\]
\ese
again for $j=1,2$, with $\omega_0(z) = 2 - (z + 1/z)$ as in \eqref{e:dLS-lineardisp}
and the same normalization conditions as before, namely,
\bse
\begin{gather} 
\label{e:dManakov-phiasymp}
\phi_n^{(j,+)}(t,z)=O(z), \quad z \to 0,	\\
\phi_n^{(j,-)}(t,z) = O(1/z), \quad z \to \infty,
\end{gather}
\ese
for $j=1,2$.
We can again convert these two RHPs into an equivalent matrix RHP by defining
\[
M_n(t,z) = \begin{pmatrix} \@I&\bPhi_n \\ \@0&\@I \end{pmatrix},
\label{e:dManakov-Mdef}
\] 
with $\bPhi_n(t,z) = \diag(\phi_n^{(1)},\phi_n^{(2)})$,
obtaining the matrix jump condition
\[
M_n^+(t,z) = M_n^-(t,z)V_n(t,z),\qquad |z|=1\,,
\]
with
\[
V_n(t,z) = \begin{pmatrix} \@I &\@U_n \\ \@0&\@I \end{pmatrix}
\label{e:dManakov-JumpMatrix}
\]
and 
\[
\@U_n(t,z) = z^{n-1}e^{-i\omega_0(z)t}\,\diag(f_1(z),f_2(z))\,.
\nonumber
\]
We then modify the jump matrix to ``nonlinearize'' the RHP by letting
\[
R_n(t,z) = V_n^\dagger(t,z^2)V_n(t,z^2) 
= \^{Z}^{n-1}e^{-i\omega(z)t\^\sigma}S(z^2),
\label{e:dManakov-Rjump}
\]
where
\[
Z = \begin{pmatrix} z\@I&\@0 \\\@0&\@I/z\end{pmatrix} \quad S(z) = \begin{pmatrix} \@I&\@F(z) \\ \@F^\dagger(z)&\@I+\@F^\dagger(z)\@F(z) \end{pmatrix}
\]
with $\@F(z)=\diag(f_1(z),f_2(z))$
and $\omega(z) = -(z - 1/z)^2/2.$
We then consider a modified matrix RHP for the $4\times4$ sectionally analytic matrix $M_n(t,z)$ still defined as in \eqref{e:dNLS-Mdef},
with the same asymptotics as before [i.e., \eqref{e:dNLS-Masymp0}],
but with the the jump condition
\[
M_n^+(t,z) = M_n^-(t,z)R_n(t,z).
\]
Again, this RHP is very similar to the RHP associated with the discrete NLS equation, 
except that every matrix is $4\times4$ instead of $2\times2$. 
A similar approach (carried out with Mathematica) can be used to show that $M_n(t,z)$ satisfies the matrix Lax pair
\bse
\label{e:matrixDNLS-LP}
\begin{gather}
M_{n+1} - \^{Z}M_n = Q_nM_nZ^{-1},  \\
\dot{M}_n + i\omega(z)[\sigma,M_n] = H_nM_n,
\end{gather}
\ese
with $\sigma = \diag(\@I,-\@I)$ and
\[
H_n(t,z) = i\sigma(Q_nZ^{-1} - Q_{n-1}Z - Q_nQ_{n-1}).
\label{e:dManakov-Hdef}
\]
Moreover,
\[
Q_n(t,z) = \begin{pmatrix} \@0&\@Q_n \\\@R_n&\@0\end{pmatrix},
\label{e:dManakov-Qdef}
\]
where $\@Q_n(t,z)$ and  $\@R_n(t,z)$ are $2\times2$ matrices,
satisfies
\[
i\sigma\dot{Q}_n + (Q_{n+1} - 2Q_{n} + Q_{n-1}) - Q_{n+1}Q_n^2 - Q_n^2Q_{n-1} = 0,
\label{e:matrixDNLS}
\] 
which is the compatibility condition of~\eqref{e:matrixDNLS-LP}.

\section{Symmetry reductions to two-component systems}
\label{s:dcoupledNLS}

The system~\eqref{e:matrixDNLS} admits self-consistent reductions that 
yield the discrete Manakov system and a novel two-component discrete system.
We discuss these next.  

\subsection{Discrete Manakov system}

Using~\eqref{e:dManakov-Qdef}, we can rewrite~\eqref{e:matrixDNLS} into a $2\times 2$ block matrix form:
\bse
\label{e:dManakov-2x2QR}
\begin{gather}
i\dot{\@Q}_n + (\@Q_{n+1}-2\@Q_n+\@Q_{n-1}) - \@Q_{n+1}\@R_n\@Q_n  - \@Q_{n}\@R_n\@Q_{n-1}= 0\,, \\
-i\dot{\@R}_n + (\@R_{n+1}-2\@R_n+\@R_{n-1}) - \@R_{n+1}\@Q_n\@R_n  - \@R_{n}\@Q_n\@R_{n-1}= 0\,.
\end{gather}
\ese
Note that in general the above system does not admit the reduction $\@R_n = \nu \@Q_n^\dagger$. 
Moreover, if we denote the entries of $\@Q_n$ and $\@R_n$ as
\[
\@Q_n = \begin{pmatrix} q_{n,1}&q_{n,2 }\\q_{n,3}&q_{n,4}\end{pmatrix},
\qquad
\@R_n = \begin{pmatrix} r_{n,1}&r_{n,3 }\\r_{n,2}&r_{n,4}\end{pmatrix},
\label{e:dManakov-QRdef}
\]
the trivial reduction $q_{n,3} = q_{n,4} = 0$ does not yield the discrete Manakov system either.
However, if we choose the entries of $\@Q_n$ and $\@R_n$ according to the symmetry reduction
\[
\@Q_n = \begin{pmatrix} q_{n,1}&q_{n,2} \\ (-1)^nr_{n,2}&(-1)^{n+1}r_{n,1}\end{pmatrix},\qquad 
\@R_n = 
\nu \@Q_n^\dagger,
\label{e:dManakov-reduction}
\]
as in~\cite{ablowitzprinaritrubatch}, we recover the discrete Manakov system:
\[
i\dot{\@q}_n + (\@q_{n+1} - 2\@q_n + \@q_{n-1}) - \nu\|\@q_n\|^2(\@q_{n+1} + \@q_{n-1}) = 0\,, 
\label{e:dManakov}
\]
where $\@q_n(t) = (q_{n,1},q_{n,2})^T$
and $\|\@q_n\|^2 = \@q_n^\dag\@q_n = q_{n,1} q_{n,1}^* + q_{n,2} q_{n,2}^*$ is the square of the Euclidean norm.
The continuum limit of~\eqref{e:dManakov} is of course the well-known Manakov system~\eqref{e:Manakov}.
Note that, by defining 
\[
v_n(t,z) = M_n(t,z)\,Z^n\,\exp[-i\omega(z)t\sigma],
\]
the Lax pair~\eqref{e:matrixDNLS-LP} then takes on a more traditional form:
\begin{gather}
v_{n+1} = X_nv_n\,,\qquad \dot{v}_n = T_nv_n\,,
\end{gather}
with 
\begin{gather}
X_n = Z + Q_n\,,\qquad 
T_n = -i\omega(z)\sigma + H_n\,.
\end{gather}
This Lax pair of the discrete Manakov system is also the same as the one found in \cite{ablowitzprinaritrubatch}.
Note, however, that while \eqref{e:dManakov} admits a straightforward continuum limit,
the same is not true for its Lax pair, due to the simultaneous presence of the factors $(-1)^n$ and $(-1)^{n+1}$ 
in \eqref{e:dManakov-reduction}.

\subsection{Novel two-component discrete and continuous systems}

Another consistent reduction of the system~\eqref{e:matrixDNLS} can be obtained by letting 
\[
\@Q_n(t) = \begin{pmatrix} q_{n,1}&q_{n,2 }\\q_{n,2}&q_{n,1}\end{pmatrix},
\label{e:d2novel-Qdef}
\]
and $\@R_n = \nu \@Q_n^\dagger$. Then~\eqref{e:dManakov-2x2QR} yields
\vspace*{-1ex}
\begin{multline}
i\dot{\@q}_n + (\@q_{n+1} - 2\@q_n + \@q_{n-1})
\\
{ }
- \nu \|\@q_n\|^2(\@q_{n+1} + \@q_{n-1}) - \nu(\@q_n^\dag\sigma_1\@q_n)\,\sigma_1(\@q_{n+1} + \@q_{n-1}) = 0\,,
\label{e:d2novel}
\end{multline}
where $\@q_n(t) = (q_{n,1},q_{n,2})^T$ as before and 
$\sigma_1$ is the first Pauli matrix, namely,
\[
\sigma_1 = \begin{pmatrix} 0 & 1 \\ 1 & 0 \end{pmatrix}.
\]
Equivalently, in component form,
\bse
\begin{gather}
\label{e:dspinor}
i\dot{q}_{n,1} + (q_{n+1,1} - 2 q_{n,1} + q_{n-1,1})
\kern20em\nonumber\\{ }
-  \nu \|\@q_n\|^2\,(q_{n+1,1} + q_{n-1,1}) 
- \nu (\@q_n^\dag\sigma_1\@q_n)\,(q_{n+1,2} + q_{n-1,2}) = 0\,, 
\\
i\dot{q}_{n,2} + (q_{n+1,2} - 2 q_{n,2} + q_{n-1,2})
\kern20em\nonumber\\{ }
-  \nu \|\@q_n\|^2\,(q_{n+1,2} + q_{n-1,2}) 
- \nu (\@q_n^\dag\sigma_1\@q_n)\,(q_{n+1,1} + q_{n-1,1}) = 0\,,
\end{gather}
\ese
where, explicitly, the scalar coupling coefficients are
\bse
\begin{gather}
\|\@q_n\|^2 = q_{n,1} q_{n,1}^* + q_{n,2} q_{n,2}^*\,,
\\
\@q_n^\dag\sigma_1\@q_n = \bar{\@q}_2^2 = q_{n,1} q_{n,2}^* + q_{n,2} q_{n,1}^*. 
\end{gather}
\ese

The system~\eqref{e:d2novel} is a discrete analogue of the two-component system that is obtained from
the full three-component spinor system [e.g., see~\eqref{e:cspinor}] by setting $q_1 = q_2$, namely, of the system
\[
i\@q_t + \@q_{xx} - 2\nu \|\@q\|^2\@q - 2 \nu(\@q^\dag\sigma_1\@q)\,\sigma_1\@q = 0\,,
\label{e:c2novel}
\]
with $\@q(x,t) = (q_1,q_2)^T$, 
to which \eqref{e:d2novel} reduces in the continuum limit.
In turn, \eqref{e:c2novel} is obtained from the $2\times2$ matrix NLS equation~\eqref{e:matrixNLS} with the symmetry reduction
\[
\@Q(x,t) = \begin{pmatrix} q_1&q_2\\q_2&q_1\end{pmatrix}.
\label{e:c2cnovel-Qdef}
\]

A natural question in light of these results is whether one can derive a discrete analogue of the full three-component spinor system~\eqref{e:cspinor}.
Note however that, if one uses for $\@Q_n$ the discrete analogue of the symmetry \eqref{e:3x3cspinor-Qdef} 
and lets $\@R_n = \nu \@Q_n^\dagger$,
one does not obtain a self-consistent reduction of the system~\eqref{e:dManakov-2x2QR}.
In order to get a three-component system, one needs to increase the dimension of the relevant matrices instead.
We do so in the next section.

\section{Novel discrete and continuous coupled integrable systems of NLS type}
\label{s:novel}

The results in section~\ref{s:dmatrixNLS} and~\ref{s:dcoupledNLS}
can be generalized to matrices with an arbitrary number of components 
in a straightforward way. 
Here, starting form three uncoupled scalar RHPs instead of two, and using similar techniques as above, 
we first derive three-component discrete and continuous integrable systems which are novel to the best of our knowledge.
We then further generalize the results to obtain novel discrete and continuous integrable systems with an arbitrarily large number of components.

\subsection{Discrete and continuous three-component integrable systems}

Consider three uncoupled sectionally analytic functions defined by 
\eqref{e:dManakov-ScalarRHP} for $j=1,2,3$, 
satisfying the same jump conditions \eqref{e:dManakov-jump} for $j=1,2,3$,
with $\omega_0(z)$ still given by \eqref{e:dLS-lineardisp}.
We can easily carry out similar calculations as in section~\ref{s:dmatrixNLS} and derive the exact system~\eqref{e:matrixDNLS}, 
except that now $\@Q_n$ and $\@R_n$ are both matrices of size $3\times 3$  instead of $2\times 2$.
Then, by letting
\[
\@Q_n = \begin{pmatrix} q_{n,1}&q_{n,2}&q_{n,3}\\q_{n,2}&q_{n,3}&q_{n,1}\\q_{n,3}&q_{n,1}&q_{n,2}\end{pmatrix},
\label{e:d3novel-Qdef}
\]
and $\@R_n = \nu \@Q_n^\dagger$, the system reduces to the three-component integrable discrete system
\begin{gather}
\dot{\@q}_n + (\@q_{n+1} - 2\@q_n + \@q_{n-1}) - \nu \|\@q_n\|^2\,(\@q_{n+1} + \@q_{n-1})
\nonumber\\{}
- \nu (\@q_n^\dag\sigma_-\@q_n)\,\sigma_+(\@q_{n+1} + \@q_{n-1})
  - \nu (\@q_n^\dag\sigma_+\@q_n)\, \sigma_-(\@q_{n+1} + \@q_{n-1}) = 0\,,
\label{e:d3novel}
\end{gather}
where $\@q_n(t) = (q_{n,1},q_{n,2},q_{n,3})^T$
and where we introduced the cyclic permutation matrices
\[
\sigma_+ = \begin{pmatrix} 0 &1 &0 \\ 0 & 0 &1 \\ 1 &0 &0 \end{pmatrix},
\qquad
\sigma_- = \sigma_+^{-1} = \begin{pmatrix} 0 &0 &1 \\ 1 & 0 &0 \\ 0 &1 &0 \end{pmatrix}.
\label{e:sigmapm}
\]
Equivalently, in component form,
\bse
\label{e:dnovel}
\begin{gather}
i\dot{q}_{n,1} + (q_{n+1,1} - 2 q_{n,1} + q_{n-1,1})
-  \nu \|\@q_n\|^2(q_{n+1,1} + q_{n-1,1}) 
\kern9em\nonumber\\{ }
- \nu (\@q_n^\dag\sigma_-\@q_n)\, (q_{n+1,2} + q_{n-1,2})
- \nu (\@q_n^\dag\sigma_+\@q_n)\,(q_{n+1,3} + q_{n-1,3}) = 0\,, 
\\
i\dot{q}_{n,2} + (q_{n+1,2} - 2 q_{n,2} + q_{n-1,2})
-  \nu (\@q_n^\dag\sigma_+\@q_n)\,(q_{n+1,1} + q_{n-1,1}) 
\kern9em\nonumber\\{ }
- \nu \|\@q_n\|^2 (q_{n+1,2} + q_{n-1,2})
- (\@q_n^\dag\sigma_-\@q_n)\, (q_{n+1,3} + q_{n-1,3}) = 0\,, 
\\
i\dot{q}_{n,3} + (q_{n+1,3} - 2 q_{n,3} + q_{n-1,3})
-  \nu (\@q_n^\dag\sigma_-\@q_n)\, (q_{n+1,1} + q_{n-1,1}) 
\kern9em\nonumber\\{ }
- \nu (\@q_n^\dag\sigma_+\@q_n)\,(q_{n+1,2} + q_{n-1,2})
- \nu \|\@q_n\|^2 (q_{n+1,3} + q_{n-1,3}) = 0\,,
\end{gather}
\ese
where, explicitly, the scalar coupling coefficients are
\bse
\begin{gather}
\|\@q_n\|^2 = q_{n,1} q_{n,1}^* + q_{n,2} q_{n,2}^* + q_{n,3} q_{n,3}^*\\
\@q_n^\dag\sigma_-\@q_n = q_{n,1} q_{n,2}^* + q_{n,2} q_{n,3}^* + q_{n,3} q_{n,1}^*\\
\@q_n^\dag\sigma_+\@q_n = q_{n,1} q_{n,3}^* + q_{n,2} q_{n,1}^* + q_{n,3} q_{n,2}^*\,.
\end{gather}
\ese
The corresponding Lax pair is still given by \eqref{e:matrixDNLS-LP},
except that now all matrices have size $6\times6$, 
with $Z = \diag(zI,I/z)$ and $\sigma = \diag(I,-I)$
[$I$ now being the $3\times3$ identity matrix],
with $Q_n(t)$ and $H_n(t,z)$ still given by \eqref{e:dManakov-Hdef} and~\eqref{e:dManakov-Qdef},
but where now $\@Q_n(t)$ and $\@R_n(t)$ are $3\times3$ matrices given by the symmetry reductions 
\eqref{e:d3novel-Qdef} and $\@R_n = \nu\@Q_n^\dag$.

Note that the system~\eqref{e:d3novel} is not a discrete analogue of the three-component spinor system
(cf.\ Appendix~\ref{s:ccoupledNLS}).
On the other hand, it is easy to see that the system~\eqref{e:d3novel} admits a continuum limit in the form
of the following three-component system:
\[
i{\@q}_t + \@q_{xx} - 2\nu \|\@q\|^2\,\@q - 2\nu (\@q^\dag\sigma_-\@q)\,\sigma_+\@q - 2\nu (\@q^\dag\sigma_+\@q)\,\sigma_-\@q = 0\,,
\label{e:c3novel}
\]
with $\@q(x,t) = (q_1,q_2,q_3)^T$ 
and with $\sigma_\pm$ given by~\eqref{e:sigmapm} as before.

The system~\eqref{e:c3novel} is not equivalent to the traditional three-component spinor system~\eqref{e:cspinor}.
(We were unable to find such a discrete analogue using the methods of this work.)
The Lax pair for~\eqref{e:c3novel} is trivially obtained from the continuum limit of \eqref{e:matrixDNLS-LP}
(unlike what happens with the discrete Manakov system),
and is simply the Lax pair of the $3\times3$ matrix NLS equation, namely \eqref{e:matrixNLS-LP},
except that all matrices are now $6\times6$,
with $\sigma$ as above, 
$Q(x,t)$ and $H(x,t,k)$ still given by \eqref{e:cManakov-Hdef} and \eqref{e:cManakov-Qdef}, and 
with the continuous analogue of the symmetry reduction~\eqref{e:d3novel-Qdef}, namely
\[
\@Q(x,t) = \begin{pmatrix} q_1&q_2&q_3\\q_2&q_3&q_1\\q_3&q_1&q_2\end{pmatrix}.
\label{e:c3cnovel-Qdef}
\]

\subsection{Discrete and continuous integrable systems with arbitrary number of components}

The two- and three-component symmetry reductions \eqref{e:d2novel-Qdef}, \eqref{e:c2cnovel-Qdef}, \eqref{e:d3novel-Qdef} and~\eqref{e:c3cnovel-Qdef} 
--- and the corresponding coupled systems of evolution equations \eqref{e:d2novel}, \eqref{e:c2novel}, \eqref{e:d3novel} and \eqref{e:c3novel} ---
admit self-consistent generalizations 
to an arbitrary number of components. 
Since the form of the continuous systems is marginally simpler than that of the corresponding discrete ones, 
for simplicity we first discuss in detail the continuous systems, and we present the corresponding discrete systems after.

The systems \eqref{e:c2novel} and~\eqref{e:c3novel} are simply the case $N=2$ and $N=3$, respectively, 
of the $N$-component system
\vspace*{-1ex}
\[
i\@q_t + \@q_{xx} -2\nu \sum_{m=0}^{N-1} (\@q^\dag\sigma_m\@q)\,\sigma_{N-m}\@q = 0\,,
\label{e:cNcomponent}
\]
where $\@q(x,t) = (q_1,\dots,q_N)^T$ and $\sigma_0,\dots,\sigma_{N-1}$ are the circular shift permutation matrices of size $N\times N$,
and $\sigma_0 = \sigma_N = I$.
That is, $\sigma_m = (\delta_{i,j-m\mod N})_{i,j=1,\dots,N}$ for all $n=0,\dots,N$,
where $\delta_{i,j}$ is the Kronecker delta.
Or, explicitly, 
\bse
\begin{gather}
\sigma_1 = \begin{pmatrix}
  0 & 1 & 0 & \cdots & 0 &0 \\
  0 & 0 & 1 & 0 & \cdots & 0 \\            
  \vdots & \vdots & \ddots & \ddots & \ddots & \vdots \\
  0 & 0 & \cdots & 0 & 1 & 0 \\
  0 & 0 & 0 & \cdots & 0 & 1 \\
  1 & 0 & 0 & 0 & \cdots & 0
\end{pmatrix},
\quad
\sigma_2 = \begin{pmatrix}
  0 & 0 & 1 & 0 & \cdots & 0 \\
  0 & 0 & 0 & 1 & \ddots & \vdots \\       
  \vdots & \vdots & \ddots & \ddots & \ddots & 0 \\
  0 & 0 & \cdots & 0 & 0 & 1 \\
  1 & 0 & 0 & \cdots & 0 & 0 \\
  0 & 1 & 0 & \cdots & 0 & 0 
\end{pmatrix}, 
\nonumber\\
\end{gather}
and so on, up to
\[
\sigma_{N-1} = \begin{pmatrix}
  0 & 0 & 0 & \cdots & 0 & 1 \\
  1 & 0 & 0 & 0 & \cdots & 0 \\
  0 & 1 & 0 & \cdots & 0 &0 \\
  0 & 0 & 1 & 0 & \cdots & 0 \\            
  \vdots & \vdots & \ddots & \ddots & \ddots & \vdots \\
  0 & 0 & \cdots & 0 & 1 & 0
\end{pmatrix},\quad
\sigma_N = I.
\]
\ese
Note that $\sigma_m = \sigma_1^m = \sigma_{N-m}^{-1}$ for $m = 0,\dots,N$.

The system \eqref{e:cNcomponent} reduces to the NLS equation~\eqref{e:NLS} for $N=1$, while
for $N=2$ and $N=3$ it yields respectively the systems \eqref{e:c2novel} and \eqref{e:c3novel}, 
as mentioned above.
Moreover, the system~\eqref{e:cNcomponent} is integrable for all $N\in\mathbb{N}$.
Indeed, it is straightforward to see that the Lax pair for the system \eqref{e:cNcomponent} is simply that of 
the $N\times N$ matrix NLS equation,
namely \eqref{e:matrixNLS} where all matrices are $2N\times2N$, 
with $\sigma = \diag(I,-I)$
[$I$ now being the $N\times N$ identity matrix],
$Q(x,t)$ and $H(x,t,k)$ still given by \eqref{e:cManakov-Hdef} and \eqref{e:cManakov-Qdef}, 
and with the symmetry reductions $\@R(x,t) = \nu \@Q^\dag(x,t)$ and
$\@Q(x,t)$ is the Hankel matrix
\[
\@Q(x,t) = \begin{pmatrix} q_1 & q_2 & q_3 & \cdots & q_{N-1} &q_N \\
  q_2 & q_3 & \cdots & q_{N-1} &q_N & q_1 \\            
  q_3 & \cdots & q_{N-1} & q_N & q_1 & q_2 \\
  \vdots & \idots & \idots & \idots & \idots & \vdots \\
  q_{N-1} & q_N & q_1 & q_2 & \cdots & q_{N-2} \\
  q_N & q_1 & q_2 & \cdots & q_{N-2} & q_{N-1}
\end{pmatrix},
\label{e:cNcomponent-Qdef}
\]
which is the $N$-component generalization of \eqref{e:c2cnovel-Qdef} and \eqref{e:c3cnovel-Qdef}.

The discrete counterpart~\eqref{e:cNcomponent} is the system
\begin{gather}
i\dot{\@q}_n + (\@q_{n+1} -2 \@q_n + \@q_{n-1}) -\nu \sum_{m=0}^{N-1} (\@q_n^\dag\sigma_m\@q_n)\,\sigma_{N-m}(\@q_{n+1} + \@q_{n-1}) = 0\,,
\label{e:dNcomponent}
\end{gather}
with $\@q_n(t) = (q_{n,1},\dots,,q_{n,N})^T$ 
and $\sigma_0,\dots,\sigma_N$ as above.
The system \eqref{e:dNcomponent} reduces to the
Ablowitz-Ladik system~\eqref{e:dNLS-AL} for $N=1$
and to \eqref{e:d2novel} and \eqref{e:d3novel} when $N=2$ and $N=3$, respectively,
as well as to \eqref{e:cNcomponent} in the continuum limit for all $N\in\mathbb{N}$.
Moreover, the system is also integrable for all $N\in\mathbb{N}$,
and its Lax pair is simply that of the $N\times N$ discrete matrix NLS equation, 
namely \eqref{e:matrixDNLS-LP} where all matrices are $2N\times2N$, 
with $Z = \diag(zI,I/z)$ and $\sigma = \diag(I,-I)$ as above, 
with $Q_n(t)$ and $H_n(t,z)$ still given by \eqref{e:dManakov-Hdef} and~\eqref{e:dManakov-Qdef},
and with the discrete analogue of the symmetry reduction~\eqref{e:cNcomponent}, namely $\@R_n(t) = \nu\@Q_n^\dag(t)$ and
$\@Q_n(t)$ is the Hankel matrix
\[
\@Q_n(t) = \begin{pmatrix} q_{n,1} & q_{n,2} & q_{n,3} & \cdots & q_{n,N-1} &q_{n,N} \\
  q_{n,2} & q_{n,3} & \cdots & q_{n,N-1} &q_{n,N} & q_{n,1} \\            
  q_{n,3} & \cdots & q_{n,N-1} & q_{n,N} & q_{n,1} & q_{n,2} \\
  \vdots & \idots & \idots & \idots & \idots & \vdots \\
  q_{n,N-1} & q_{n,N} & q_{n,1} & q_{n,2} & \cdots & q_{n,N-2} \\
  q_{n,N} & q_{n,1} & q_{n,2} & \cdots & q_{n,N-2} & q_{n,N-1}
\end{pmatrix},
\label{e:dNcomponent-Qdef}
\]

Each member of the above infinite families of discrete and continuous integrable systems 
is novel to the best of our knowledge,
except obviously in the special case $N=1$, 
in which case one simply has the NLS equation and its integrable discretization.

Of course one could object that the systems \eqref{e:cNcomponent} and \eqref{e:dNcomponent} are simply reductions of the continuous 
and discrete $N\times N$ matrix NLS equations.
However, we point out that apparently these reductions, and the vector systems themselves, were not previously known.
Moreover, one should realize that, in order for a full $N\times N$ system of equations to admit a self-consistent symmetry reduction to an $N$-component system,
a total of $N(N-1)$ different constraints must be satisfied
[namely, the equality of the evolution equations for all the various matrix entries associated to the same vector component].
Therefore, the existence of compatible reductions is nontrivial, in general.
Nonetheless, the symmetry \eqref{e:cNcomponent-Qdef} and its discrete counterpart~\eqref{e:dNcomponent-Qdef}
are such all that those constraints are satisfied identically.
Another way to look at the situation is to note that the symmetry reduction~\eqref{e:cNcomponent-Qdef} and its discrete counterpart~\eqref{e:dNcomponent-Qdef}
define invariant manifolds in the set of solutions of the continuous and discrete matrix NLS equations, respectively.
In other words, the symmetry reductions are preserved by the time evolution: 
if one assigns initial conditions that satisfy these symmetries at $t=0$ (i.e., if the initial conditions lie on the manifold), 
the symmetries will remain valid at all times $t\ne0$
(i.e., the state of the system remains on the manifold at all times, both positive and negative).

\section{Conclusions}
\label{s:remarks}

In summary, we have presented the discrete analogue of the dressing method and used it to derive a variety of discrete integrable systems,
together with their continuum limit.
We end this work with some final remarks.

1. We should point out that the approach in section~\ref{s:dLS} is trivially generalized to arbitrary linear discrete evolutions,
whose Lax pair was obtained in \cite{BiondiniWang2010}.
It should also be clear that, if one is interested in continuum limits, it is straightforward to repeat all of the calculations 
when one inserts a lattice spacing constant $h$ in the equations.
Indeed, this is precisely how the system 
\eqref{e:c2novel} was obtained from \eqref{e:d2novel} in section~\ref{s:dcoupledNLS}
and the system \eqref{e:c3novel} from \eqref{e:d3novel} in section~\ref{s:dcoupledNLS}.

2. We should also note that, while the method is algorithmic, it does not always yield nontrivial results.
In the continuous case this happens, for example, when the dispersion relation is linear, like for the one-directional wave equation $q_t + q_x = 0$.
In this case, if one ``dresses'' the corresponding RHP in the same way as in \cite{pinotsis2007}, one does not obtain a nonlinear integrable system.
The situation is different in the discrete case, however.
Indeed, in Appendix~\ref{s:mKdV}, starting from the discrete counterpart of the above equation, namely a discrete one-directional wave equation, 
we derive a discrete integrable system of equations together with its Lax pair.
We then discuss the compatible reductions admitted by this system, including that to a discrete integrable modified KdV equation.

3. We reiterate that, 
while some of the integrable nonlinear systems of equations derived here were known,
the method can also yield novel systems.
In the continuous case, this was recently demonstrated in~\cite{BiondiniWang2015}, where novel coupled nonlinear systems of PDEs
describing resonant wave interactions were derived using the continuous version of the dressing method.
Here we used similar techniques to derive two infinite families of discrete and continuous integrable systems
which are novel to the best of our knowledge.

4. On the other hand, we were unable to successfully use the method to produce integrable discretizations of the two- and three-wave interaction equations.
It is therefore an interesting question whether even further novel discrete coupled integrable systems can be derived using the method presented in this work.

5. From a physical point of view,
the main difference between the coupled NLS equations and the novel systems obtained in this work
is that the nonlinear interaction terms in coupled NLS equations only describe self-phase and cross-phase modulation, whereas those in the 
novel systems derived here also include nontrivial four-wave mixing interactions.

6. In light of the above observation, an interesting and important practical question is therefore 
whether some of these systems can be derived from first principles in some of the physical contexts where the NLS equation arises.
For example, the two- and three-component discrete integrable systems \eqref{e:d2novel} and \eqref{e:d3novel}
and their continuum limits \eqref{e:c2novel} and \eqref{e:c3novel}
are related to the dynamical equations that describe coupled systems of Bose-Einstein condensates,
and are therefore likely to be physically relevant.
As another example, the two- and three-component systems describe the evolution of interacting quasi-monochromatic 
optical fields under certain kinds of nonlinear couplings.  Are there practical situations where these couplings arise?

7. It is also worthwhile to note that, in addition to the reductions 
$\@R_n(t) = \nu\@Q_n(t)$ (in the discrete case) and $\@R(x,t) = \nu\@Q(x,t)$ (in the continuous case)
all the ``QR'' systems derived in this work also admit the additional self-consistent reductions
$\@R_n(t) = \nu\@Q_{-n}(t)$ and $\@R(x,t) = \nu\@Q(-x,t)$, respectively.
These reductions immediately yield discrete and continuous coupled nonlocal systems which are generalizations of the 
nonlocal NLS equations recently studied in \cite{AM1,AM2,AM3}.

8. Finally, from a mathematical point of view, 
the novel families of discrete and continuous systems derived in this work open up a number of interesting questions, 
relating to whether the integrable structure (e.g., hierarchies, conservation laws, symmetries, inverse scattering transform, etc.)
and the behavior of the solution of these systems differ significantly from those for the standard $N$-component vector NLS equation.

We hope that the results of this work and the above questions will stimulate further work on this topics.


\section*{Acknowledgments}

We thank Dimitris Pinotsis for helpful interactions on an early portion of this work 
and Mark Ablowitz, Athanassios Fokas and Barbara Prinari for insighftul discussions.
This work was partially supported by the National Science Foundation under grant numbers  
DMS-1614623 and 
DMS-1615524.


\section*{Appendix}
\setcounter{section}0
\setcounter{equation}0
\def\thesection{\Alph{section}}
\renewcommand{\theequation}{\Alph{section}.\arabic{equation}}
\numberwithin{equation}{section}

In section~\ref{s:cNLS}
we review the continuous version of the dressing method by showing how it can be used it to derive the NLS equation
together with their Lax pair.
In section~\ref{s:ccoupledNLS} we generalize the method to derive the matrix NLS equation and its Lax pair.
Finally, in section~\ref{s:mKdV} we use the discrete version of the method to derive a discrete modifier KdV equation 
starting from a discrete one-directional wave equation.

\section{Linear and nonlinear Schr\"{o}dinger equations via dressing}
\label{s:cNLS}

Starting form the RHP associated with the linear Schr\"{o}dinger equation, 
we now show how to construct an appropriate matrix RHP and derive the NLS equation
\[
iq_t + q_{xx} - 2\nu|q|^2q = 0\,,
\label{e:NLS}
\]
together with its Lax pair. 
We refer the reader to~\cite{fokas2008, pinotsis2007} for further details.


Consider the scalar RHP associated with the linear Schr\"{o}dinger equation defined by the jump condition
\bse
\label{e:LS-RHP}
\[
\mu^+ - \mu^- = e^{i\theta(x,t,k)}f(k) \qquad k\in \Real\,,
\label{e:LS-jump}
\] 
for the sectionally analytic function
\[
\mu(x,t,k) = \begin{cases} \mu^+(x,t,k), & k\in\mathbb{C}^+, \,\\ \mu^-(x,t,k), & k\in\mathbb{C}^-,\end{cases}
\label{e:LS-Mdef}
\]
with $\theta(x,t,k)=kx - k^2t$ and the normalization condition
\[
 \mu = O(1/k) \qquad k\to \infty.
 \label{e:LS-masymp1}
\]
\ese
This problem can be trivially converted into a matrix RHP by letting
\[
M(x,t,k) = \begin{pmatrix} 1&\mu(x,t,k) \\0&1\end{pmatrix},
\label{e:NLS-Mdef}
\] 
obtaining the jump condition
\[
M^+(x,t,k) = M^-(x,t,k)V(x,t,k),\qquad k\in \Real\,,
\label{e:NLS-jump}
\]
for the sectionally analytic $2\times2$ matrix
\[
M(x,t,k)=
\begin{cases}
    M^+(x,t,k)\,, \qquad k\in\Complex^+,  \\
    M^-(x,t,k)\,, \qquad k\in\Complex^-,  \\
  \end{cases}
\label{e:NLS-Mdef2x2}
\]
with jump matrix
\[
V(x,t,k) =  \begin{pmatrix} 1&e^{i\theta(x,t,k)}f(k) \\0&1\end{pmatrix},
\label{e:NLS-Jdef}
\]
and normalization condition
\[
M(x,t,k) = I + O(1/k), \qquad k\to\infty\,,
\label{e:NLS-RHPnormalization}
\]
where $I$ denotes the identity matrix of appropriate size (in this case $2\times2$).
The key to nonlinearize the problem 
is to modify the jump matrix in~\eqref{e:NLS-jump}.
In particular, following \cite{fokasgelfand1994}, we replace $V(x,t,k)$ in~\eqref{e:NLS-jump} with:
\[
R(x,t,k) = V^\dagger(x,t,2k)V(x,t,2k) 
= e^{i(kx-2k^2t)\^\sigma_3}S(2k)
\label{e:NLS-Rdef}
\]
where
\[
\sigma_3 = \begin{pmatrix} 1&0 \\0&-1\end{pmatrix}\,,\qquad 
S(k) = \begin{pmatrix} 1&f(k) \\f^*(k)&1+|f(k)|^2\end{pmatrix},
\label{e:NLS-sigmadef}
\]

We therefore consider a modified matrix RHP for the sectionally analytic matrix $M(x,t,k)$ in \eqref{e:NLS-Mdef2x2},
with jump condition
\[
M^+(x,t,k) = M^-(x,t,k)R(x,t,k),\qquad k \in \Real\,,
\label{e:NLS-Rjump}
\]
and the same normalization condition~\eqref{e:NLS-RHPnormalization}. 
Below we show that any solution $M(x,t,k)$ of this new RHP satisfies the matrix Lax pair
\bse
\label{e:NLS-LP}
\begin{gather}
    M_x - ik[\sigma_3,M] = QM,  \\
    M_t + 2ik^2[\sigma_3,M] = HM,
\end{gather}
\ese
with 
\[
H = i\sigma_3(Q_x - Q^2) - 2kQ\,,
\label{e:NLS-Hdef}
\]
and
\[
Q(x,t) = -i\lim_{k \to \infty} [\sigma_3,kM]=\begin{pmatrix} 0&q \\r&0\end{pmatrix}\,.
\]
Moreover, the matrix $Q(x,t)$
satisfies
\[
i\sigma_3Q_t + Q_{xx} -2Q^3 = 0,
\label{e:2x2NLS}
\] 
which is the compatibility condition of~\eqref{e:NLS-LP}.
In turn, note that equation~\eqref{e:2x2NLS} admits the self-consistent reduction 
$r(x,t) = \nu q^*(x,t)$, with $\nu=\pm1$.
which yields the NLS equation~\eqref{e:NLS}.
Note that the resulting Lax pair is equivalent to the one satisfied the Jost eigenfunctions \cite{ablowitzprinaritrubatch},
as can be seen by performing the transformation $v(x,t,k) = \exp[ik\sigma_3x-2ik^2\sigma_3t]$.

The way to construct the two linear differential operators $L$ and $N$ 
follows similar ideas as before. 
That is, the operators are partially determined by the jump condition of the RHP and partially by the required asymptotic behavior. 
%
%
%
First we construct the operator $L$ which will yield the scattering problem by defining
\[
LM = M_x - ik[\sigma_3,M] - QM.
\label{e:NLS-Ldef}
\]
It is straightforward to see that $LM$ satisfies the same jump condition as $M$, namely:
\[
LM^+ = (LM^-)R.
\label{e:NLS-LMjump}
\]
We next check that $LM = O(1/k)$ as $k\to\infty$.
To do so, using the normalization condition~\eqref{e:NLS-RHPnormalization} we can write an asymptotic expansion for $M(x,t,k)$ as:
\[
M(x,t,k) = I + \frac{M_1(x,t)}{k} + \frac{M_2(x,t)}{k^2} +O(1/k^3) \qquad k \to \infty.
\label{e:NLS-Masymp2}
\]
Substituting~\eqref{e:NLS-Masymp2} into~\eqref{e:NLS-Ldef} we have
\[
LM = \lbrace -i[\sigma_3,M_1] - Q\rbrace + O(1/k) \qquad k \to \infty.
\label{e:NLS-Lasymp}
\]
Since we want $LM$ to be $O(1/k)$, we need the $O(1)$ term to be identically zero, which implies
\[
Q = -i[\sigma_3,M_1] = -i\lim_{k \to \infty} [\sigma_3,kM].
\label{e:NLS-Qdef}
\]
Then we can conclude that 
\[
LM = M_x - ik[\sigma_3,M] - QM = 0,
\nonumber
\]
which gives us the first half of~\eqref{e:NLS-LP}.


We now construct the operator $N$, which yields the time evolution equation.
Define
\[
NM = M_t + 2ik^2[\sigma_3,M] - kA(x,t)M - B(x,t)M,
\label{e:NLS-Ndef}
\]
where $A$ and $B$ are to be determined and do not depend on $k$.
Then, similarly as before,
\[
NM^+ = (NM^-)R.
\label{e:NLS-Njump}
\]
Next, substituting~\eqref{e:NLS-Masymp2} into~\eqref{e:NLS-Ndef} yields
\[
NM = \lbrace 2i[\sigma_3,M_1] - A\rbrace k + \lbrace 2i[\sigma_3,M_2] - AM_1 - B\rbrace  + O(1/k)
\label{e:NLS-Nasymp}
\]
as $k\to\infty$.
Since we want $ NM = O(1/k)$ as well, we need both the $O(k)$ and $O(1)$ terms to be zero, which implies
\bse
\begin{gather}
\label{e:NLS-Adef}
A = 2i[\sigma_3,M_1] = -2Q\,, \\
\label{e:NLS-Bdef}
B = 2i[\sigma_3,M_2] - AM_1 = 2i[\sigma_3,M_2] + 2QM_1\,.
\end{gather}
\ese
To get the information of $M_2$, we look at the $O(1/k)$ term of $LM$
\[
LM = \lbrace M_{1x} - i[\sigma_3,M_2] - QM_1\rbrace/k + O(1/k^2) \qquad k \to \infty.
\]
Since $LM$ vanishes identically, the $O(1/k)$ term then leads to
\[
i[\sigma_3,M_2] + QM_1 = M_{1x}.
\label{e:NLS-M1d}
\]
Thus,
\[
B = 2i[\sigma_3,M_2] + 2QM_1 = 2M_{1x}.
\]
We now split $M_1$ into its diagonal and off-diagonal parts $M_1^d$ and $M_1^o$, respectively. From~\eqref{e:NLS-Qdef} we get
$M_1^o = (i/2)\,\sigma_3Q$, while 
the diagonal part of equation~\eqref{e:NLS-M1d} gives us
$M^d_{1x} = - (i/2)\,\sigma_3Q^2$.
Combining these results we then have
\[
B = 2M_{1x} = 2(-\frac{i}{2}\sigma_3Q^2 + \frac{i}{2}\sigma_3Q_x) = i\sigma_3(Q_x - Q^2)
\]
and
\[
NM = M_t + 2ik^2[\sigma_3,M] - kAM - BM = 0,
\nonumber
\]
which gives us the second half of~\eqref{e:NLS-LP}.

\section{Matrix NLS equation via dressing}
\label{s:ccoupledNLS}

We now generalize the method of section~\ref{s:cNLS} to coupled systems.
Namely, starting from two uncoupled linear Schr\"{o}dinger equations and their associated RHPs, 
we construct an appropriate matrix RHP and use it we derive the matrix NLS equations. 
Under proper reductions, we then obtain the Manakov system and the spinor system.
The process is the same as before apart from the change in matrix dimensionality, and therefore we omit the details.
As in section~\ref{s:dmatrixNLS} we will use boldface letters to denote $2\times 2$ matrices.


We start with two uncoupled scarlar RHPs~\eqref{e:LS-RHP}, that is, we consider two sectionally analytic functions
\bse
\label{e:cManakov-ScalarRHP} 
\[
\mu_j(x,t,k) = \begin{cases} \mu_j^+(x,t,k), & k\in\mathbb{C}^+, \,\\ \mu_j^-(x,t,k), & k\in\mathbb{C}^-,\end{cases}
\]
for $j=1,2$, satisfying jump conditions
\[
\mu_j^+-\mu_j^-=e^{i\theta_j(x,t,k)}f_j(k) \qquad k\in \mathbb{R},\quad j= 1,2,
\label{e:cManakov-jump} 
\]
with $\theta_j(x,t,k)=kx-k^2t\,$ and the same normalization conditions
\[
\mu_j(x,t,k) = O(1/k) \qquad k\to\infty\, \qquad j=1,2. 
\label{e:cManakov-normalization}
\]
\ese 
These two RHPs can be converted into an equivalent matrix RHP by letting
\[
M(x,t,k) = \begin{pmatrix} \@I&\bmu \\ \@0&\@I \end{pmatrix},
\label{e:cManakov-Mdef}
\] 
with $\@I$ and $\@0$ the $2\times2$ identity and zero matrix respectively, and $\bmu(x,t,k) = \diag(\mu_1,\mu_2)$,
obtaining again the matrix jump condition \eqref{e:NLS-jump}, except that $M(x,t,k)$ and $V(x,t,k)$ are now $4\times4$ matrices 
instead of $2\times2$.  
In particular,
\[
V(x,t,k) = \begin{pmatrix} \@I &\@U \\ \@0&\@I \end{pmatrix}
\label{e:cManakov-JumpMatrix}
\]
and 
\[
\@U(x,t,k) = \diag(e^{i\theta_1(x,t,k)}f_1(k),\,e^{i\theta_2(x,t,k)}f_2(k))~.
\nonumber
\]
The corresponding normalization condition is the same as \eqref{e:NLS-RHPnormalization}, except that 
$I$ is now the 4$\times$4 identity matrix.

Again we modify the jump matrix by letting
\[
R(x,t,k) = V^\dagger(x,t,2k)V(x,t,2k) 
= e^{i(kx-2k^2t)\^\sigma}S(2k)
\label{e:cManakov-Rdef}
\]
with
\[
S(k) = \begin{pmatrix} \@I&\@F(k) \\ \@F^\dagger(k)&\@I+\@F^\dagger(k)\@F(k) \end{pmatrix}
\]
and $\@F(k)=\diag(f_1(k),f_2(k))$.

Now we have a modified matrix RHP for a $4\times4$ matrix $M(x,t,k)$ defined by \eqref{e:NLS-Mdef2x2},
with normalization~\eqref{e:NLS-RHPnormalization}
and jump condition
\[
M^+(x,t,k) = M^-(x,t,k)R(x,t,k).
\label{e:cManakov-RJump}
\]
It is not hard to see that this RHP is very similar to the RHP associated with the scalar NLS equation, 
except that every matrix is $4\times4$ instead of $2\times2$. 
A similar approach can be used to show that $M(x,t,k)$ satisfies the matrix Lax pair
\bse
\label{e:matrixNLS-LP}
\begin{gather}
    M_x - ik[\sigma,M] = QM,  \\
    M_t + 2ik^2[\sigma,M] = HM,
\end{gather}
\ese
with 
\[
H = i\sigma(Q_x - Q^2) - 2kQ.
\label{e:cManakov-Hdef}
\]
Moreover,
\[
Q = -i\lim_{k \to \infty} [\sigma,kM]=\begin{pmatrix} \@0&\@Q \\\@R&\@0\end{pmatrix},
\label{e:cManakov-Qdef}
\]
where $\@Q(x,t)$ and  $\@R(x,t)$ are $2\times2$ matrices,
satisfies
\[
i\sigma Q_t + Q_{xx} -2Q^3 = 0,
\label{e:cManakov}
\] 
which is the compatibility condition of~\eqref{e:matrixNLS-LP}.


Using~\eqref{e:cManakov-Qdef}, we can rewrite~\eqref{e:cManakov} into a $2\times 2$ block matrix form:
\bse
\label{e:cManakov-2x2QR}
\begin{gather}
i\@Q_t + \@Q_{xx} - 2\@Q\@R\@Q = 0\,, \\
-i\@R_t + \@R_{xx} - 2\@R\@Q\@R = 0\,.
\end{gather}
\ese
The above system admits the reduction $\@R = \nu \@Q^\dagger$, which yields the matrix NLS equation
\[
i\@Q_t + \@Q_{xx} - 2 \nu \@Q\@Q^\dagger\@Q = 0\,. 
\label{e:matrixNLS}
\]


The $2\times2$ matrix NLS system~\eqref{e:matrixNLS} admits various self-consistent reductions.
Denoting the entries of $\@Q$ as
\[
\@Q = \begin{pmatrix} q_1&q_2 \\q_3&q_4\end{pmatrix},
\label{e:cManakov-QRdef}
\]
under further reduction $q_3 = q_4 = 0$,~\eqref{e:matrixNLS} yields the Manakov system:
\[
i\@q_t + \@q_{xx} - 2 \nu \|\@q\|^2\@q = 0, 
\label{e:Manakov}
\]
for $\@q(x,t) = (q_1,q_2)^T$, and where $\|\@q\|^2 = |q_1|^2+|q_2|^2$ is the standard Euclidean norm.
Note that by letting
$v(x,t,k) = M(x,t,k)\,\exp[ik\sigma x - 2ik^2\sigma t]$,
the Lax pair~\eqref{e:matrixNLS-LP} can be transformed into the more traditional form 
\begin{gather}
\label{e:cManakov-mLP}
v_x = Xv\,,\qquad v_t = Tv\,,
\end{gather}
with 
\begin{gather}
X = -ik\sigma + Q\,,\qquad 
T = 2ik^2 \sigma + H\,.
\end{gather}
Note however that this is still not the traditional Lax pair for the Manakov system, since it has size $4\times 4$,
whereas the usual Lax pair for~\eqref{e:Manakov} is $3\times3$ \cite{manakov,ablowitzprinaritrubatch}.
However, under the reduction $q_3=q_4=0$, 
the second row and the second column of both matrices $Q$ and $H$ are zero. 
Therefore,  the corresponding $4\times 4$ Lax pair can be reduced to a $3\times 3$ one by simply deleting the second row and the second column, which then reduces~\eqref{e:cManakov-mLP} to the usual $3\times 3$ Lax pair for the Manakov system.

Another self-consistent reduction of the matrix NLS equation \eqref{e:matrixNLS} can be obtained by taking
\[
\@Q = \begin{pmatrix} q_1&q_3 \\q_3&q_2\end{pmatrix}.
\label{e:3x3cspinor-Qdef}
\]
Then~\eqref{e:matrixNLS} yields the three-component system
\bse
\label{e:cspinor}
\begin{align}
&i q_{1,t} + q_{1,xx} - 2 \nu (|q_1|^2+|q_3|^2) q_1 - 2 \nu (q_1 q_3^* + q_2^* q_3) q_3 = 0, \\
&i q_{2,t} + q_{2,xx} - 2 \nu (|q_2|^2+|q_3|^2) q_2 - 2 \nu (q_2 q_3^* + q_1^* q_3) q_3 = 0, \\
&i q_{3,t} + q_{3,xx} - 2 \nu (|q_1|^2+|q_2|^2+|q_3|^2) q_3 - 2 \nu q_1 q_2 q_3^* = 0.
\end{align}
\ese
The above system can be easily transferred into equivalent forms of the spinor system 
(e.g., such as the one proposed in \cite{spinor1})
via simple rescalings.

\section{Discrete modified KdV equation via dressing}
\label{s:mKdV}


Consider the following discrete one-directional wave equation:
\[
\dot q_n + (q_{n+1}-q_{n-1}) = 0\,.
\label{e:LinearEqn}
\]
It was shown in~\cite{BiondiniWang2010} that \eqref{e:LinearEqn} admits the Lax pair
\bse
\label{e:linearLP}
\begin{gather}
\phi_{n+1} - z \phi_n = q_n,  \\
\dot \phi_n + i\tilde{\omega}_0(z)\phi_n = -(q_n + q_{n-1}/z),
\end{gather}
\ese
where the linear dispersion relation is now
\[
\tilde{\omega}_0(z) = -i (z - 1/z).
\label{e:omega0Def}
\]
The spectral analysis of~\eqref{e:linearLP} leads to the same scalar jump condition as before, namely \eqref{e:dLS-jump},
for the sectionally analytic function $\phi_n(t,z)$ in \eqref{e:dLS-phidef},
except that $\omega_0(z)$ is now replaced by $\tilde\omega_0(z)$.
The normalization conditions for $\phi_n(t,z)$ are the same as before, namely, \eqref{e:dLS-phiasymp}.

The above scalar RHP problem can be converted into an equivalent $2\times 2$ matrix RHP in a similar way as in section~\ref{s:dNLS} by introducing the matrix $M_n(t,z)$ as in \eqref{e:dLS-matrixdef},
obtaining again the matrix jump condition \eqref{e:dLS-matrixjump} across the countour $|z|=1$, 
with jump matrix $V_n(t,z)$ as in \eqref{e:dLS-jumpmatrix}.
One then modifies the jump matrix by replacing $V_n(t,z)$ with $R_n(t,z)$ defined by \eqref{e:dNLS-Rdef} as before,
except that now $\tilde{\omega}(z) = -i(z^2 - 1/z^2)/2$.
As in section~\ref{s:dNLS} we then have
a modified matrix RHP for the sectionally analytic function $M_n(t,z)$ defined in \eqref{e:dNLS-Mdef}, 
with asymptotic behavior specified by~\eqref{e:dNLS-Masymp0} and the same 
jump condition~\eqref{e:dNLS-jump}, 
except that once again $\tilde{\omega}(z) = -i(z^2 - 1/z^2)/2$.

The same approach as in section~\ref{s:dNLS} can then be carried out algorithmically to show that $M_n(t,z)$ satisfies the 
same matrix Lax pair, namely, \eqref{e:dNLS-LP}, except that $\omega(z)$ is replaced with $\tilde\omega(z)$, and 
\[
H_n = Q_n Q_{n-1} - Q_n Z^{-1} - Q_{n-1} Z .
\]
Moreover, the matrix \eqref{e:dNLS-Qdef}
satisfies
\[
\label{e:matrixeqn}
\dot Q_n + Q_{n+1} - Q_{n-1} - Q^2_n (Q_{n+1} - Q_{n-1}) = 0,
\]
which is the compatibility condition of the matrix Lax pair. 
Finally, imposing the symmetry $r_n = q_n$,~\eqref{e:matrixeqn} reduces to 
\[
\label{e:scalareqn}
\dot q_n + q_{n+1} - q_{n-1} - q_n^2 (q_{n+1} - q_{n-1}) = 0\,,
\]
which is an integrable discretization of the modified KdV equations~\cite{tsuchida1998}.
(A different integrable discretization of the modified KdV equation, corresponding to the
discrete dispersion relation $\tilde\omega(z) = z^4/2 - z^2 + 1/z^2 - 1/2z^4$, 
had already been found in \cite{ablowitzladik}.)


\begin{thebibliography}{Oi}
\frenchspacing
\small

\bibitem{AC1991}
M. J. Ablowitz and P. A. Clarkson, 
\textit{Solitons, Nonlinear Evolution Equations and Inverse Scattering}
(Cambridge University Press, Cambridge 1991)

\bibitem{ablowitzfokas}
M. J. Ablowitz and A. S. Fokas, 
\textit{Complex Variables: Introduction and Applications} 
(Cambridge University Press, 2003)

\bibitem{akns1974}
M. J. Ablowitz, D. J. Kaup, A. C. Newell and H. Segur, 
\textit{The inverse scattering transform - Fourier analysis for nonlinear problems}, 
Stud. Appl. Math. \textbf{53}, 249-315, (1974)

\bibitem{ablowitzladik}
M. J. Ablowitz and J. F. Ladik,
\textit{Nonlinear differential-difference equations and Fourier analysis}, 
J. Math. Phys. \textbf{17}, 1011-1018, (1976)

\bibitem{ablowitzhalburdherbst}
M. J. Ablowitz, R. Halburd and B. Herbst,
\textit{On the extension of the Painlevé property to difference equations},
Nonlinearity \textbf{13}, 889 (2000)

\bibitem{AM1}
M. J. Ablowitz and Ziad H. Musslimani, 
\textit{Integrable nonlocal nonlinear Schr\"odinger equation},
Phys. Rev. Lett. \textbf{110}, 064105 (2013)

\bibitem{AM2}
M. J. Ablowitz and Z. Musslimani,
\textit{Integrable discrete PT symmetric model},
Phys. Rev. E \textbf{90}, 032912 (2014)

\bibitem{AM3}
M. J. Ablowitz and Z. Musslimani,
\textit{Inverse scattering transform for the integrable nonlocal nonlinear Schr\"odinger equation},
Nonlinearity \textbf{29}, 915 (2016)

\bibitem{ablowitzprinaritrubatch}
M. J. Ablowitz, B. Prinari, and A. D. Trubatch, 
\textit{Discrete and continuous nonlinear Schr\"odinger systems} 
(Cambridge University Press, 2004)

\bibitem{ablowitzsegur1981}
M. J. Ablowitz and H. Segur, 
\textit{Solitons and the inverse scattering transform} (SIAM, Philadelphia, 1981)

\bibitem{BBEIM1994}
E. D. Belokolos, A. I. Bobenko, V. Z. Enol’skii, A. R. Its and V. B. Matveev, 
\textit{Algebro-Geometric Approach to Nonlinear Integrable Equations} 
(Springer, Berlin 1994)

\bibitem{BiondiniHwang2008}
G. Biondini and G. Hwang,
\textit{Initial-boundary value problems for discrete evolution equations: discrete linear Schr\"odinger and integrable discrete nonlinear Schr\"odinger equations},
Inv. Probl. \textbf{24}, 1-44, (2008)

\bibitem{BiondiniWang2010}
G. Biondini and D. Wang,
\textit{Initial-boundary-value problems for discrete linear evolution equations},
IMA J. Appl. Math. \textbf{75}, 968-997, (2010)

\bibitem{BiondiniWang2015}
G. Biondini and Q. Wang,
\textit{Novel systems of resonant wave interactions},
J. Phys. A \textbf{48}, 225203 (2015)

\bibitem{bobenkoseiler}
A. I. Bobenko and R. Seiler, Eds.,
\textit{Discrete integrable geometry and physics}
(Clarendon Press, Oxford, 1999)

\bibitem{bruschi1}
M. Bruschi, O. Ragnisco, P. M. Santini and T. Gui-Zhang,
\textit{Integrable symplectic maps},
Phys. D  \textbf{49}, 273--294 (1991)

\bibitem{bruschi2}
M. Bruschi, P. M. Santini and O. Ragnisco, 
\textit{Integrable cellular automata}, 
Phys. Lett. A \textbf{169}, 151--160 (1992)

\bibitem{doliwa}
A Doliwa and P M Santini, 
\textit{Multidimensional quadrilateral lattices are integrable}, 
Phys. Lett. A \textbf{233}, 365--372 (1997)

\bibitem{doliwa2}
A Doliwa, P M Santini and M Ma\~{n}as,
\textit{Transformations of quadrilateral lattices},
J. Math. Phys. \textbf{41}, 944 (2000)

\bibitem{FT1987}
L. D. Faddeev and L. A. Takhtajan, 
\textit{Hamiltonian Methods in the Theory of Solitons}
(Springer, Berlin, 1987)

\bibitem{flaschka}
H. Flaschka,
\textit{On the Toda lattice. II. Existence of integrals},
Phys. Rev. B \textbf{9}, 1924--1925 (1974)

\bibitem{flaschka2}
H. Flaschka,
\textit{On the Toda lattice. II: Inverse-scattering solution},
Progr. Theor. Phys. \textbf{51}, 703--716 (1974)

\bibitem{fokas2008}
A. S. Fokas, 
\textit{A unified approach to boundary value problems} 
(SIAM, Philadelphia, 2008)

\bibitem{fokasgelfand1994}
A. S. Fokas and I. M. Gelfand, \textit{Integrability of linear and nonlinear evolution equations and the associated nonlinear Fourier transforms}, 
Lett. Math. Phys. \textbf{32}, 189--210 (1994)

\bibitem{fokaszakharov}
A. S. Fokas and V. E. Zakharov,
\textit{The dressing method and nonlocal Riemann-Hilbert problems},
J. Nonlinear Sci. \textbf{2}, 109--134 (1992)

\bibitem{Gakhov}
F. D. Gakhov,
\textit{Boundary value problems}
(Pergamon Press, 1966)


\bibitem{GH1990}
F. Gesztesy and H. Holden, 
\textit{Soliton Equations and Their Algebro-geometric Solutions} 
(Cambridge University Press, Cambridge, 1990)

\bibitem{goktashereman}
U. G\"oktas and W. Hereman,
\textit{Computation of conservation laws for nonlinear lattices},
Phys. D \textbf{123}, 425--436 (1998)

\bibitem{grammaticos}
B Grammaticos, A Ramani, J Satsuma, R Willox, AS Carstea,
\textit{Reductions of integrable lattices},
J. Nonlin. Math. Phys. \textbf{12}, 363--371 (2005)

\bibitem{halburd}
R. Halburd and R. J. Korhonen,
\textit{Finite-order meromorphic solutions and the discrete Painlev\'e equations},
Proc. London Math. Soc. \textbf{94}, 443-474 (2006)

\bibitem{hietarintajoshinijhoff}
J. Hietarinta, N. Joshi and F Nijhoff,
\textit{Discrete Systems and Integrability}
(Cambridge University Press, 2016)

\bibitem{spinor1}
J. Ieda, T. Miyakawa and M. Wadati,
\textit{Exact analysis of soliton dynamics in spinor Bose-Einstein condensates},
Phys. Rev. Lett. \textbf{93}, 194102 (2004)

\bibitem{kajiwarasatsuma}
K Kajiwara and J Satsuma,
\textit{The conserved quantities and symmetries of the two‐dimensional Toda lattice hierarchy},
J. Math. Phys. \textbf{32}, 506 (1991)

\bibitem{kakei}
S Kakei, JJC Nimmo, R Willox,
\textit{Yang-Baxter maps and the discrete KP hierarchy},
Glasgow Math. J. \textbf{51}(A), 107--119 (2009)

\bibitem{Kevrekidis}
P. G. Kevrekidis, D. J. Frantzeskakis, and R. Carretero-Gonz\'alez, Eds., 
\textit{Emergent Nonlinear Phenomena in Bose-Einstein Condensates} 
(Springer, New York, 2008)

\bibitem{leviwinternitz}
D. Levi and P. Winternitz,
\textit{Continuous symmetries of discrete equations},
Phys. Lett. A \textbf{152}, 335--338 (1991)
Physics Letters A

\bibitem{manakov}
S. V. Manakov,
\textit{On the theory of two-dimensional stationary self-focusing of electromagnetic waves},
Sov. Phys. JETP \textbf{38}, 248-253, (1974)

\bibitem{NMPZ1984}
S. Novikov, S. V. Manakov, L. P. Pitaevskii and V. E. Zakharov, \textit{Theory of solitons: the inverse scattering method} (Plenum, New York, 1984)
 
\bibitem{ohta}
Y Ohta, R Hirota, S Tsujimoto and T Imai,
\textit{Casorati and Discrete Gram Type Determinant Representations of Solutions to the Discrete KP Hierarchy},
Author information
J. Phys. Soc. Japan \textbf{62}, 1872--1886  (1993)

\bibitem{ohtahirota}
Y Ohta and R Hirota,
\textit{A discrete KdV equation and its Casorati determinant solution},
J. Phys. Soc. Japan \textbf{60}, 2095  (1991)

\bibitem{pinotsis2007}
D. A. Pinotsis, 
\textit{The Riemann--Hilbert formalism for certain linear and nonlinear integrable PDEs}, 
J. Nonlinear Math. Phys. \textbf{14}, 466--485 (2007)

\bibitem{Pitaevskii}
P. Pitaevskii and S. Stringari, 
\textit{Bose-Einstein condensation}, 
Clarendon Press, Oxford (2003)

\bibitem{santini}
P. M. Santini, 
\textit{Multiscale expansions of difference equations in the small lattice spacing regime, and a vicinity and integrability test. I},
J. Phys. A \textbf{43}, 045209  (2010)

\bibitem{spinor2}
P. Szankowski, M. Trippenbach, E. Infeld and G Rowlands,
\textit{A class of compact entities in three component Bose--Einstein condensates}, 
Phys. Rev. A \textbf{83}, 013626 (2011)

\bibitem{toda}
M. Toda,
\textit{Vibration of a chain with a non-linear interaction}, 
J. Phys. Soc. Japan \textbf{22}, 431--436 (1967)

\bibitem{TrogdonOlver}
T. Trogdon and S. Olver,
\textit{Riemann-Hilbert problems, Their Numerical Solution and the Computation of Nonlinear Special Functions}
(SIAM, Philadelphia, 2016)

\bibitem{tsuchida1998}
T. Tsuchida, H. Ujino and M. Wadati, 
\textit{Integrable semi-discretization of the coupled modified KdV equations}, 
J. Math. Phys. \textbf{39}, 4785 (1998)

\bibitem{spinor3}
M. Wadati and N. Tsuchida,
\textit{Wave propagations in the F=1 spinor Bose--Einstein condensates},
J. Phys. Soc. Jpn. \textbf{75}, 014301 (2006)

\bibitem{Yang}
J. Yang,
\textit{Nonlinear waves in integrable and non-integrable systems}
(SIAM, Philadelphia, 2010)


\bibitem{zakharovshabat1}
V. E. Zakharov and A. B. Shabat, 
\textit{A scheme for integrating nonlinear equations of mathematical physics by the method of the inverse scattering problem, Part I},
Funct. Anal. Appl. \textbf{8}, 43--53 (1974)

\bibitem{zakharovshabat2}
V. E. Zakharov and A. B. Shabat, 
\textit{A scheme for integrating nonlinear equations of mathematical physics by the method of the inverse scattering problem, Part II},
Funct. Anal. Appl. \textbf{31}, 13--22 (1979)

\bibitem{zhou}
X. Zhou,
\textit{The Riemann-Hilbert problem and inverse scattering},
SIAM J. Math. Anal. \textbf{20}, 966--986 (1989)

\end{thebibliography}
\end{document}